\documentclass[10pt,letterpaper]{article}
\usepackage{opex3}
\usepackage{amsmath}
\usepackage{amsfonts}
\usepackage{amssymb}
\usepackage{amsthm}
\usepackage{graphicx}
\usepackage{float}
\usepackage{epstopdf}
\usepackage{caption}
\usepackage{subcaption}

\begin{document}

\title{Optimized multifrequency light collection by adaptive self-ordering of scatterers in optical resonators}

\author{Valentin Torggler and Helmut Ritsch}

\address{Institute for Theoretical Physics, University of Innsbruck\\Technikerstrasse 25/2, A-6020 Innsbruck, Austria}

\email{valentin.torggler@student.uibk.ac.at}


\begin{abstract}
Mobile light scatterers in a high-Q optical cavity transversely illuminated by laser light close to a cavity resonance form ordered patterns, which maximize light scattering into the cavity and induce optical self-trapping. We show that a generalized form of such crystallization dynamics appears in multicolored pump fields with several cavity modes. Here the particles arrange in spatial patterns maximizing total light collection into the resonator. For changing input frequencies and strengths the particles dynamically adapt to the current illumination. Interestingly the system keeps some memory on past configurations, so that a later renewed application of the same pattern exhibits faster adaptation towards optimal collective scattering. In a noisy environment particles explore larger regions of configuration space spending most of the time close to optimum scattering configurations. This adaptive self-ordering dynamics should be implementable in a wide range of systems ranging from cold atoms in multimode cavities or nano-fiber traps to molecules or mobile nano-particles within an optical resonator.
\end{abstract}


\bibliographystyle{osajnl}
\bibliography{references_selforg}

\section{Introduction}
Polarizable particles in an optical resonator which are coherently illuminated from the side at sufficient intensity can undergo a phase transition from homogeneous to crystalline order accompanied by superradiant light scattering into the cavity \cite{domokos2002collective,black2003observation,arnold2012self}. This transition can be understood from a simultaneous maximization of collective scattering of pump light into the cavity mode and the depth of the optical trap created via interference of pump and cavity light \cite{griesser2010vlasov,ritsch2013cold}. In a monochromatic plane wave geometry the particles form a Bragg-like grating structure, which optimally couples cavity and pump wave so that the intensity of the scattered light grows with the square of the particle number. In a lossy cavity the ordering process is dissipative and in the long time limit light is superradiantly scattered into the cavity. For varying color and geometry of the illumination the particles evolve towards configurations with maximal collective scattering rendering the system an adaptive self-optimizing light collection device.

Here we study this self-ordering dynamics with several light frequencies applied simultaneously, pushing the particles towards competing order. As generic example we use a multicolor configuration with several light frequencies tuned closely to different cavity modes. As the longitudinal cavity modes form an equidistant comb of distinct resonances, a corresponding illumination is implementable with standard comb technology at comparable technical complexity to the single mode case. For sufficiently distinct pump frequencies, light scattering between different modes can be neglected and the computational complexity of the model grows linearly with the number of modes. Besides studying the multitude of stationary solutions of the coupled atom-field dynamics we will also numerically study the time evolution for larger particle ensembles and illumination frequencies including momentum diffusion terms. An even more complex dynamics arises for cavities with almost degenerate mode families as e.g. in a confocal cavity, which already exhibits a very rich structure for a single pump frequency \cite{domokos2002dissipative, gopalakrishnan2009emergent}. Note that even for particles in free space collective light scattering can induce some spatial bunching \cite{dholakia2010colloquium,bowman2013optical, schmittberger2012free,chang2013self,labeyrie2014optomechanical} and complex motional dynamics via collective light scattering are also predicted for multi-frequency optical lattices \cite{ostermann2014scattering}.

After a short presentation of the semi-classical dynamical model in section 2, we will study the forces and stationary states in generic two and three particle configurations in section 3. Typical scenarios of dynamical evolution trajectories in particle-field phase space of growing complexity are studied in section 4, which is followed by long term studies of many-particle dynamics including memory effects.
\section{Model}

\begin{figure}
\centering
\includegraphics[width=5cm]{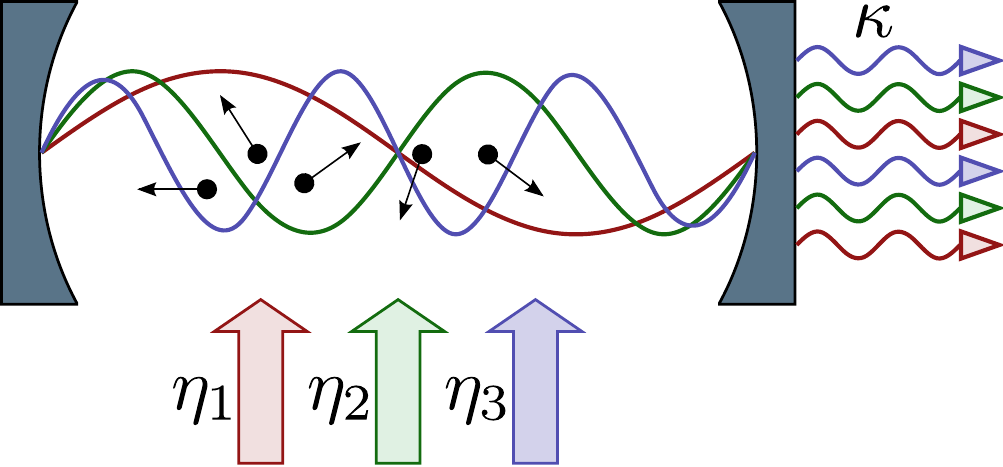}
\caption{Scheme of particle ensemble moving inside a lossy optical resonator characterized by the decay rate $\kappa$ illuminated by various laser beams with pump strengths $\eta_1$, $\eta_2$ and $\eta_3$.}
\label{model}
\end{figure}

We consider $N$ identical polarizable point particles of mass $m$ representing e.g. individual atoms, molecules or nano-particles trapped within an optical resonator supporting a large number of modes of similar finesse with wave numbers $k_n := nk$ and frequencies $\omega_n$ (see Fig. \ref{model}). The particles are illuminated transversely by lasers with frequencies $\omega_{p,n}$, each closely tuned to one of the cavity modes but sufficiently detuned from internal optical excitations to ensure linear polarizability and negligible spontaneous emission. Each particle scatters light in and out of the $n$-th cavity mode with phase and coupling strength $\eta_n \sin(k_n x_j)$  depending on its position $x_j$. For computational simplicity we restrict the particles' motion along the cavity axis but 2D and 3D traps should lead to essentially similar physics. Following standard adiabatic elimination procedures the coherent dynamics of the system can be described by the effective Hamiltonian \cite{ritsch2013cold,domokos2003mechanical}
\begin{equation}
H = \sum_{j=1}^N \frac{p_j^2}{2m} - \hbar \sum_{n \in I} \left[ \left( \delta_{c,n} - U_{0,n} \sum_{j=1}^N \sin^2(k_n x_j) \right) a_n^\dagger a_n +  \eta_n \sum_{j=1}^N \sin(k_n x_j) (a_n + a_n^\dagger)\right].
\end{equation}
Here $\eta_n$ is the effective pump strength, $U_{0,n}$ the light shift per photon and $\delta_{c,n} := \omega_{p,n} - \omega_{n}$ the detuning between laser and cavity mode frequency. For nanoparticles $U_{0,n}$ is directly proportional to the particle's polarizability \cite{nimmrichter2010master}. $x_j$ and $p_j$ denote position and momentum of the $j$-th particle, while $a_n$ and $a_n^\dagger$ are the bosonic annihilation and creation operators of the $n$-th cavity mode field. The illumination pattern is specified by choosing a subset  $I \subset \mathbb{N}$ of modes with pump amplitudes $\eta_n$. Including dissipation via cavity photon loss requires to solve the master equation $\dot{\rho} = - \frac{i}{\hbar} [H,\rho] + \mathcal{L} \rho $ with Liouvillian damping operators $\mathcal{L} \rho := \sum_n \kappa_n (2 a_n \rho a_n^\dagger - a_n^\dagger a_n \rho - \rho a_n^\dagger a_n)$.

While the full quantum model exhibits intriguing physical behavior as quantum phase transitions even for a single frequency \cite{gopalakrishnan2009emergent,baumann2010dicke}, the enlarged case of two modes and few particles already touches the limits of current numerical computability \cite{kramer2014self}. Here we focus on the essential physics of crystallization and collective light scattering for many particles and modes, which forces us to simplifications. Hence we treat particle motion classically and assume coherent states with complex amplitudes $\alpha_n$ for the cavity modes. Fortunately, this approximation works well in related treatments of cavity cooling \cite{domokos2003mechanical} and self-ordering. The corresponding semi-classical equations for the coupled particle-mode dynamics can be written as \cite{domokos2003mechanical}
\begin{subequations}
 \begin{flalign}
  \dot{x}_j &= \frac{p_j}{m},  \;\;\;
  \dot{p}_j = - \hbar \sum_{n \in I} k_n \left( U_{0,n} |\alpha_n|^2 \sin (2 k_n x_j) + \eta_n (\alpha_n + \alpha_n^*) \cos (k_n x_j) \right)\label{time_evo_2}\\
  \dot{\alpha}_n &= i \left( \delta_{c,n} - U_{0,n} \sum_{j=1}^N \sin^2 (k_n x_j) \right) \alpha_n - \kappa_n \alpha_n - i \eta_n \sum_{j=1}^N \sin(k_n x_j) + \xi_n.\label{time_evo_3}
 \end{flalign}
\label{time_evo}
\end{subequations}
The Langevin noise term $\xi_n$ and the weak frequency-dependence of $U_0$, $\delta_c$ and $\kappa$ will be mostly neglected in the following, but
these equations, obtained in similar form previously \cite{domokos2002dissipative}, still contain the essence of multicolor self-ordering. Generalizations to include spontaneous emission or collisions are straightforward in principle but much harder to analyze in practice.


\section{Self-ordered states and light scattering of few particles in a multicolored field}

Before starting an intensive numerical analysis let us first examine a few simple but instructive few particle cases where we look for stationary states. In the bad cavity limit, where cavity losses happen on a shorter time-scale compared to particle motion, the field adiabatically follows the particle positions and can be expressed in the form
\begin{equation}
 \alpha_n (x_1,...,x_N) = \eta_n \frac{\sum_j \sin (k_n x_j)}{\delta_c - U_0 \sum_j \sin^2 (k_n x_j) + i \kappa},
\end{equation}
where we neglect the $n$-dependence of $U_0$, $\delta_c$ and $\kappa$. This immediately shows that a homogeneous particle distribution possesses a close to zero field as the sum in the numerator will vanish. Only an ordered configuration will show significant scattering with an amplitude proportional to the particle number in the best case, when all sines in the sum are of same sign and order 1. Such a configuration is connected to a particular wave vector $k_n$ but one can envisage particle distributions which are close to optimal for several distinct frequencies. Note that the denominator shows resonant scattering enhancement if $\delta_c$ is suitably shifted from the bare cavity resonance $\delta_c=0$ to induce a deep optical potential and stable particle order.\\
In the bad cavity limit the force on the $j$-th particle (\ref{time_evo_2}) thus is effectively a function $F_j(x_1,\dots,x_N)$ of the particle positions. Hence, when we determine the points in configuration space where all $F_j$'s are vanishing we obtain the equilibrium points of the system. To determine their stability in a strongly damped case, where $\dot{x}_j \propto F_j(x_1,\dots,x_N)$, a stability criterion which only involves checking the sign of the real part of the eigenvalues of the Jacobian of the vector containing the $F_j$'s is employed.

\subsection{Two particles}
\begin{figure}[h]
\begin{subfigure}[b]{4.33cm}
\includegraphics[height=3.93cm]{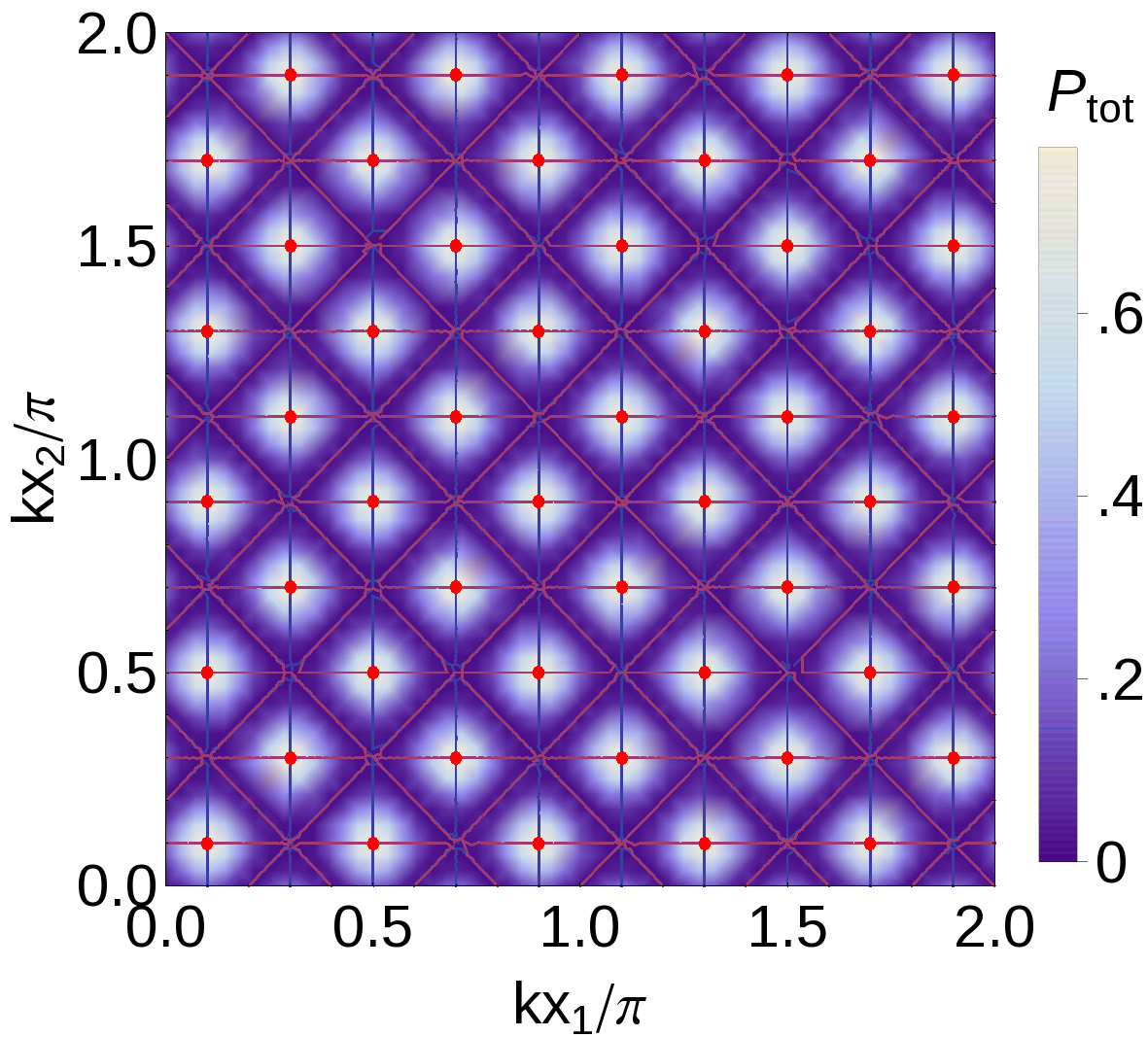}
\caption{}\label{photonnumber2-left}
\end{subfigure}
\begin{subfigure}[b]{4.33cm}
\includegraphics[height=3.93cm]{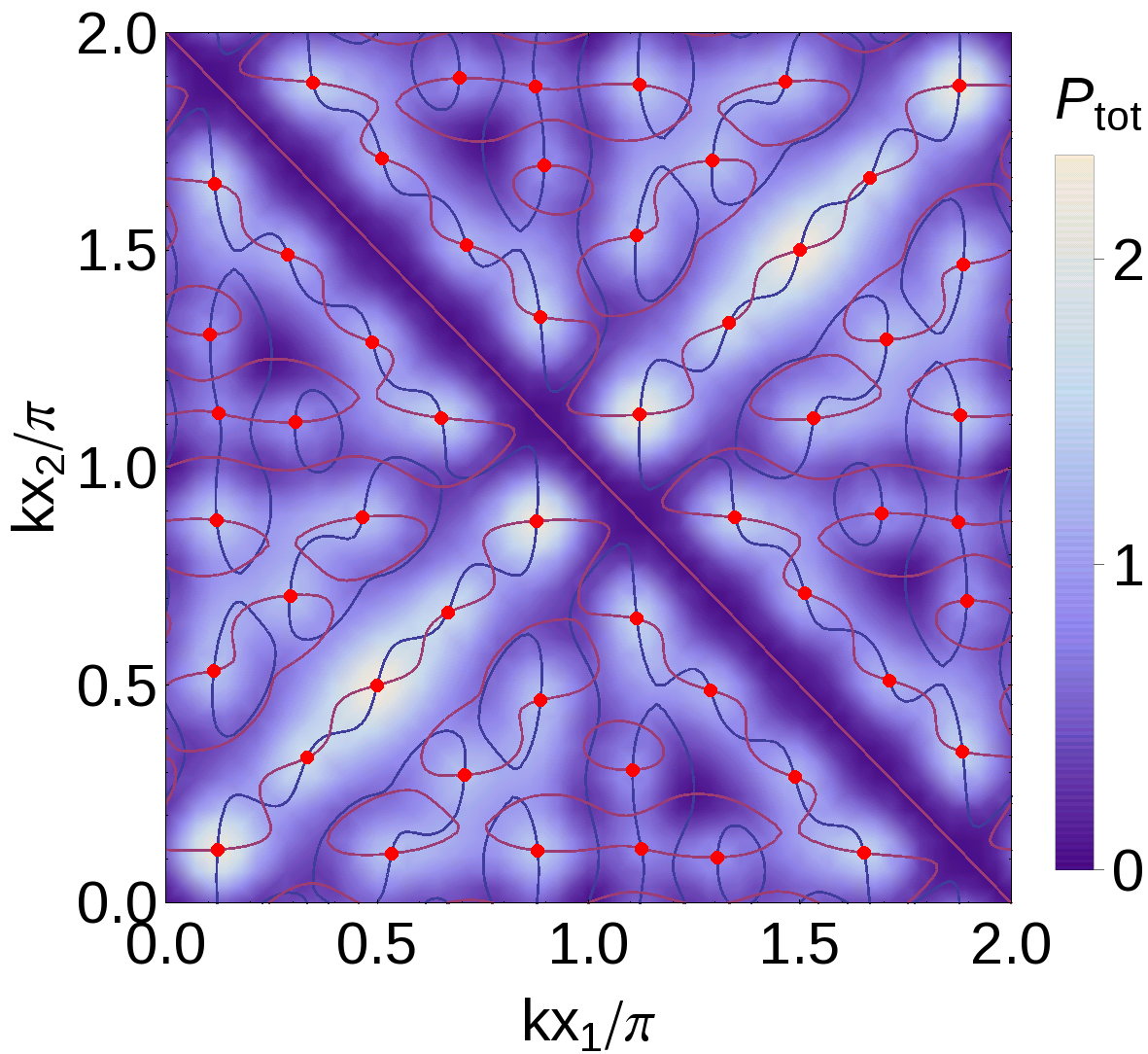}
\caption{}\label{photonnumber2-center}
\end{subfigure}
\begin{subfigure}[b]{4.33cm}
\includegraphics[height=3.93cm]{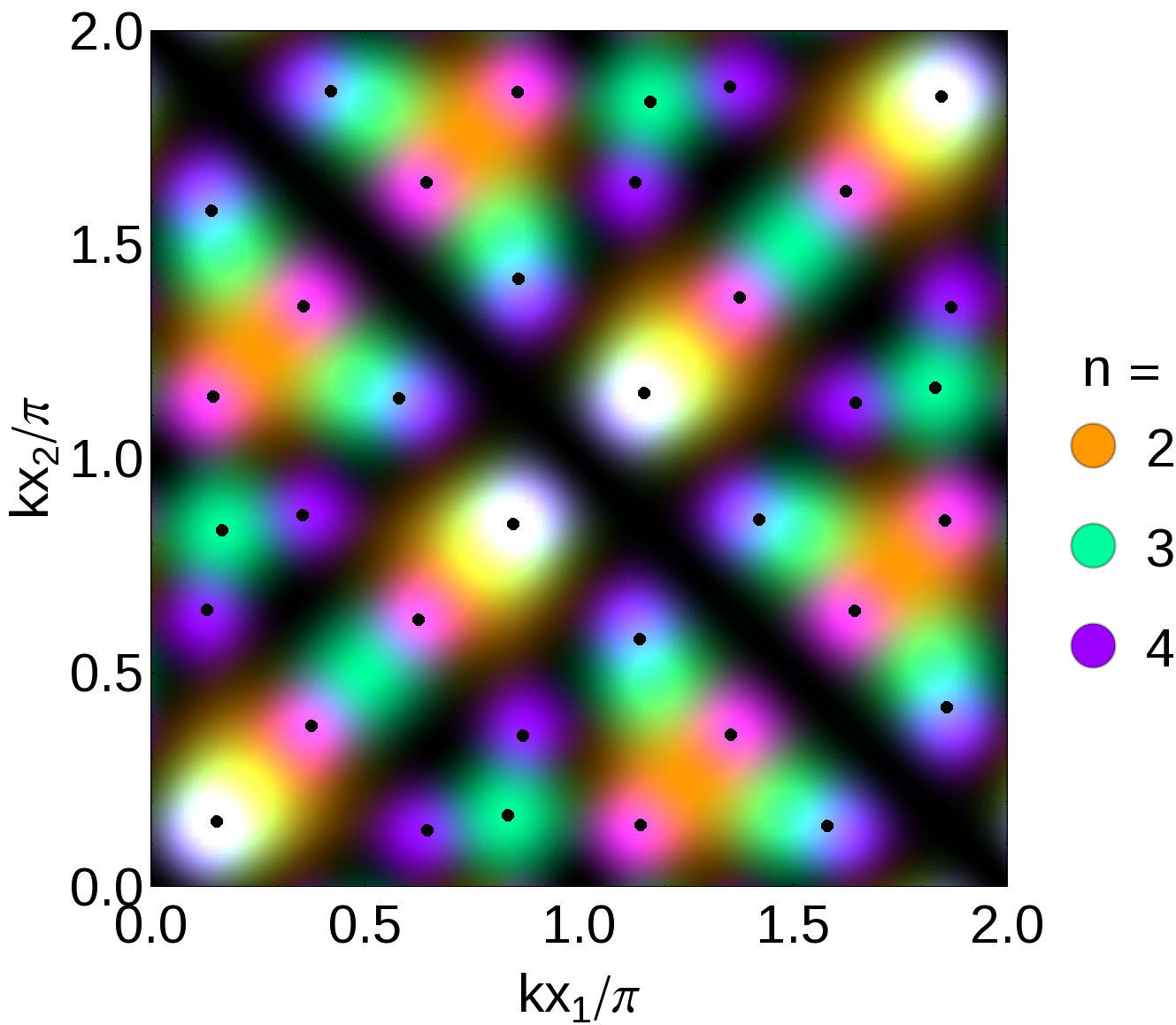}
\caption{}\label{photonnumber2-right}
\end{subfigure}
\caption{Scattered light intensity and stable points (red dots) for two atoms as function of their positions within one wavelength of the fundamental mode for different illuminations. In (a) and (b), the density plot visualizes the cavity light intensity $P_{tot}$ and the contours are zero-force lines for each atom. In (c), the individual mode intensities are color-coded leading to mixing when modes overlap. The atoms are pumped close to $\omega_n$, $n \in \{1,2,3,4,5\}$ with amplitudes (a) $\eta_n = \eta (0,0,0,0,1)_n$, (b) $\eta_n = \eta (1,0,1,1,1)_n$ and (c) $\eta_n = \eta (0,1,1,1,0)_n$. Here we have set $\eta = 5 \kappa / 8$, $NU_0 = -\kappa/10$, $\delta_c = NU_0 - \kappa$.}
\label{photonnumber2}
\end{figure}

As first non-trivial example we consider only two particles subject to transverse red-detuned pumping with multiple frequencies, where we first choose low order cavity modes to graphically better exhibit the physics. Depending on their positions they scatter a different fraction of the various pump fields into the cavity. For the known case of a single pump frequency as shown in Fig. \ref{photonnumber2-left}, we find a periodic pattern of equally strong scattering when both particles sit on various field anti-nodes with the same phase. Adding extra pump lasers close to other longitudinal modes only slightly shifts the stable equilibrium points with respect to the single pump frequency configuration and creates some extra equilibria (see Fig. \ref{photonnumber2-center}). However, the total amount of scattered light $P_{tot} = \sum_n |\alpha_n|^2$ at each stable point now strongly varies, which implies different local trap depths. In a statistical equilibrium distribution one will then find the particles more likely in deeper potential wells. As these are naturally associated with stronger light scattering, the system adapts towards optimal light collection. Note that strong scattering more likely occurs along the diagonal, where both particles occupy the same well and scatter with exactly the same phase.

Clearly, the optimum scattering positions for the different frequencies do not coincide, as demonstrated in Fig. \ref{photonnumber2-right}, where we color-code three different scattering intensities by different colors. For most particle positions one or two colors dominate but there are a few spots with nearly equal intensity scattering yielding `white' scattered light.

\subsection{Three particles}
\begin{figure}[h]
\begin{subfigure}[b]{4.15cm}
\includegraphics[height=4.5cm]{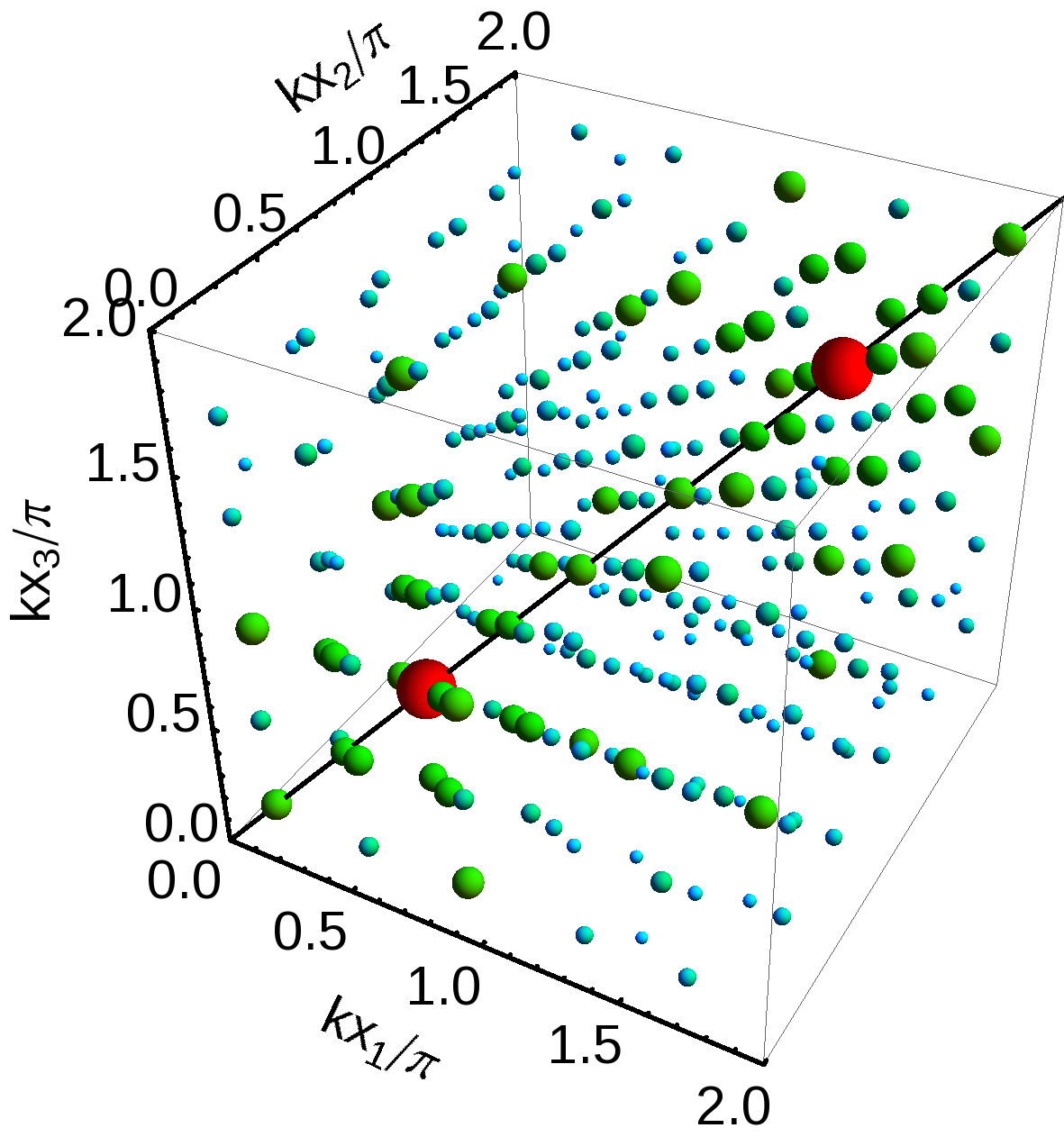}
\caption{}\label{3dspheres-left}
\end{subfigure}
\begin{subfigure}[b]{4.8cm}
\includegraphics[height=4.5cm]{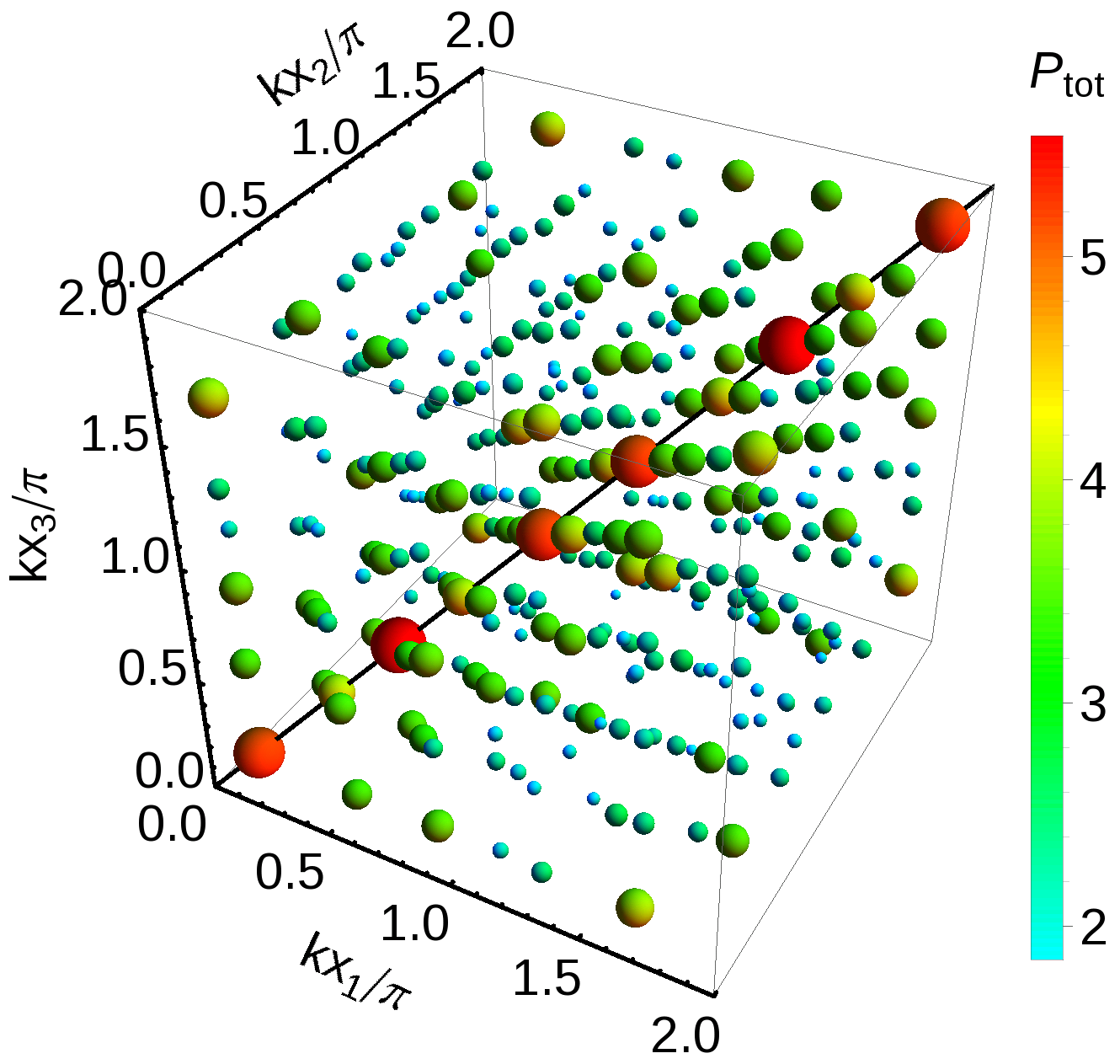}
\caption{}\label{3dspheres-center}
\end{subfigure}
\begin{subfigure}[b]{4.15cm}
\includegraphics[height=4.5cm]{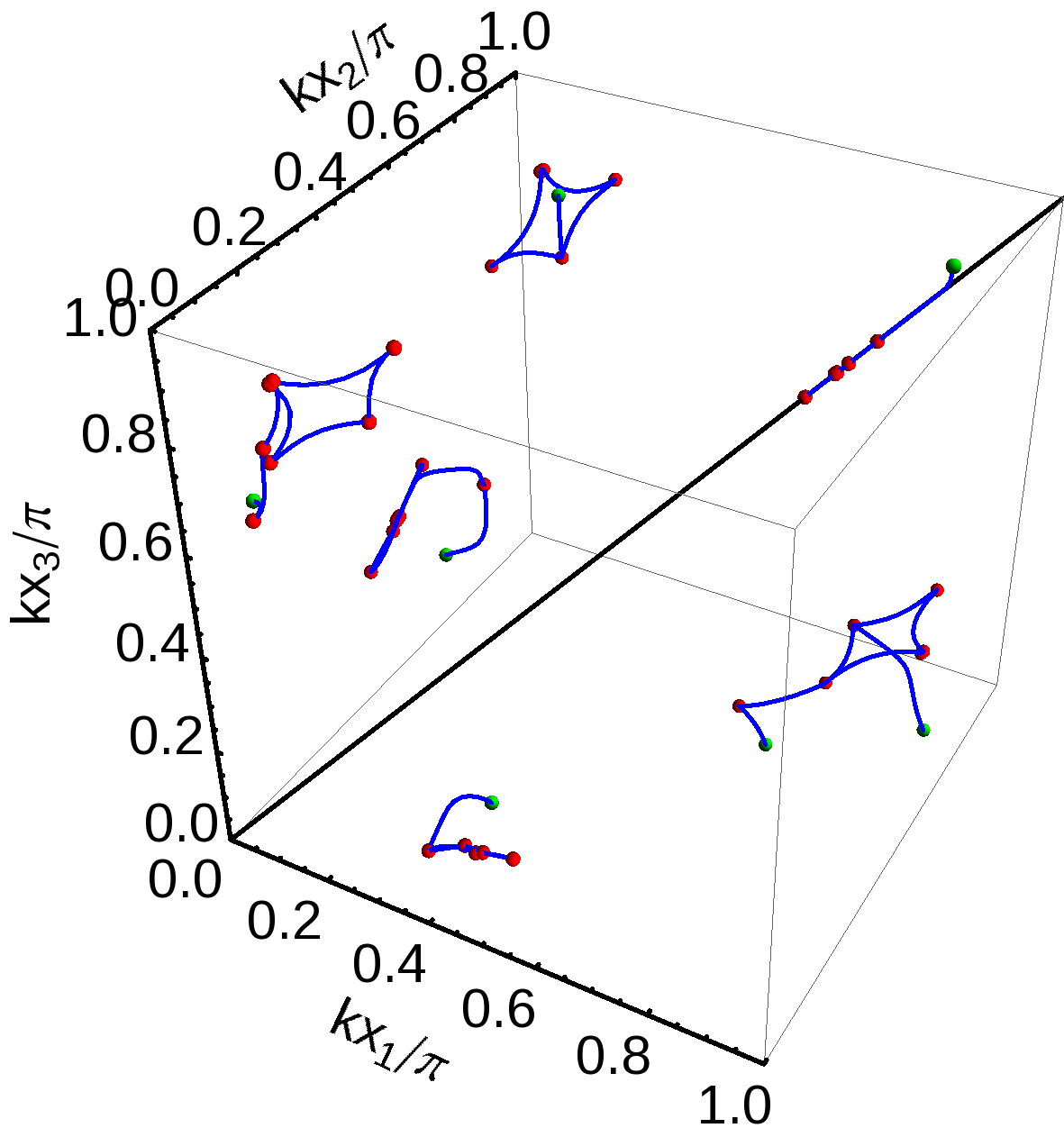}
\caption{}\label{3dspheres-right}
\end{subfigure}
\caption{Stable equilibrium configurations for three particles represented by spheres, whose size and color encode the amount of scattered light $P_{tot}$. Illumination is set to (a) $\eta_n = \eta (1,0,1,0,1)_n$ and (b) $\eta_n = \eta (1,0,1,1,1)_n$ (same as \ref{photonnumber2-center}) with other parameters as in Fig. \ref{photonnumber2}. Subplot (c) shows trajectories for periodically time-varying illumination for different initial positions, when $\eta_n/\eta$ periodically cycles through $(1,0,1,0,0,0,1)$, $(0,1,1,0,1,1,0)$, $(0,0,1,0,1,0,0)$, $(0,1,1,1,1,1,0)$, $(1,1,1,1,0,1,0)$. The illumination changes after the system reaches a stable point (red dots). Parameters in (c) are: $\eta=\kappa/5$, $NU_0=-\kappa$ and $\delta_c=NU_0/2-2\kappa$.}
\label{3dspheres}
\end{figure}

The more complex case of three particles and several frequencies invokes a configuration space of a cube of length $\lambda = 2 \pi / k$ . The distribution of the stable equilibrium points for two different illumination conditions is depicted in Figs. \ref{3dspheres-left} and \ref{3dspheres-center} by small spheres, where size and color of the spheres represent the total amount of scattered light at the corresponding stable point. Again the points with maximum scattering are on the diagonal (i.e. all three particles are in the same well). Note that the number of stable points is much larger than for two particles; it strongly increases with particle number.

\section{Adaptive dynamics of the coupled atom field system}
\label{subsec:time-dep-ill}

So far we studied light fields and forces at fixed particle positions, where points of vanishing force give equilibrium positions. Equations (\ref{time_evo}), however, also describe dynamical properties of the system. As known from the single mode case the  delayed field response during the self-ordering for negative detuning induces friction (cavity cooling) so that the system reaches a steady state \cite{niedenzu2011kinetic}. Hence in the following numerical simulations we add an effective linear friction by which we represent such cavity cooling as well as other friction terms.

\subsection{Dynamics for strongly damped particle motion}
Let us study the coupled particle-field-dynamics in a simplified form first, where the friction force is large so that the particles quickly relax to a stationary velocity for a given light force (overdamped case). The light fields in turn will continuously adapt to the current particle positions and the coupled system evolves to a close-by equilibrium configuration exhibiting a local light scattering maximum similar to the case of a self-consistent optical lattice \cite{asboth2008optomechanical}. When the illumination condition changes the particles will evolve towards a new equilibrium point, which better adapts to the momentary pump field configuration. 
We visualize this dynamics by periodically repeating a series of different pump light patterns. In this case after some initial position changes, the particles find a suitable closed loop in configuration space and periodically follow the illumination sequence. In each step they quickly arrange to a local optimum configuration to maximize combined light scattering. Due to the complexity of the optical potential landscape, for different initial conditions a multitude of such loops are attained for the same illumination sequence. These can be distinguished by a characteristic corresponding output intensity sequence. Typical sample paths for a specific illumination sequence with different initial positions are shown in Fig. \ref{3dspheres-right} nicely demonstrating this effect.

\subsection{Dynamics with noise forces on the particles}
\begin{figure}[h]
\centering
\includegraphics[width=10cm]{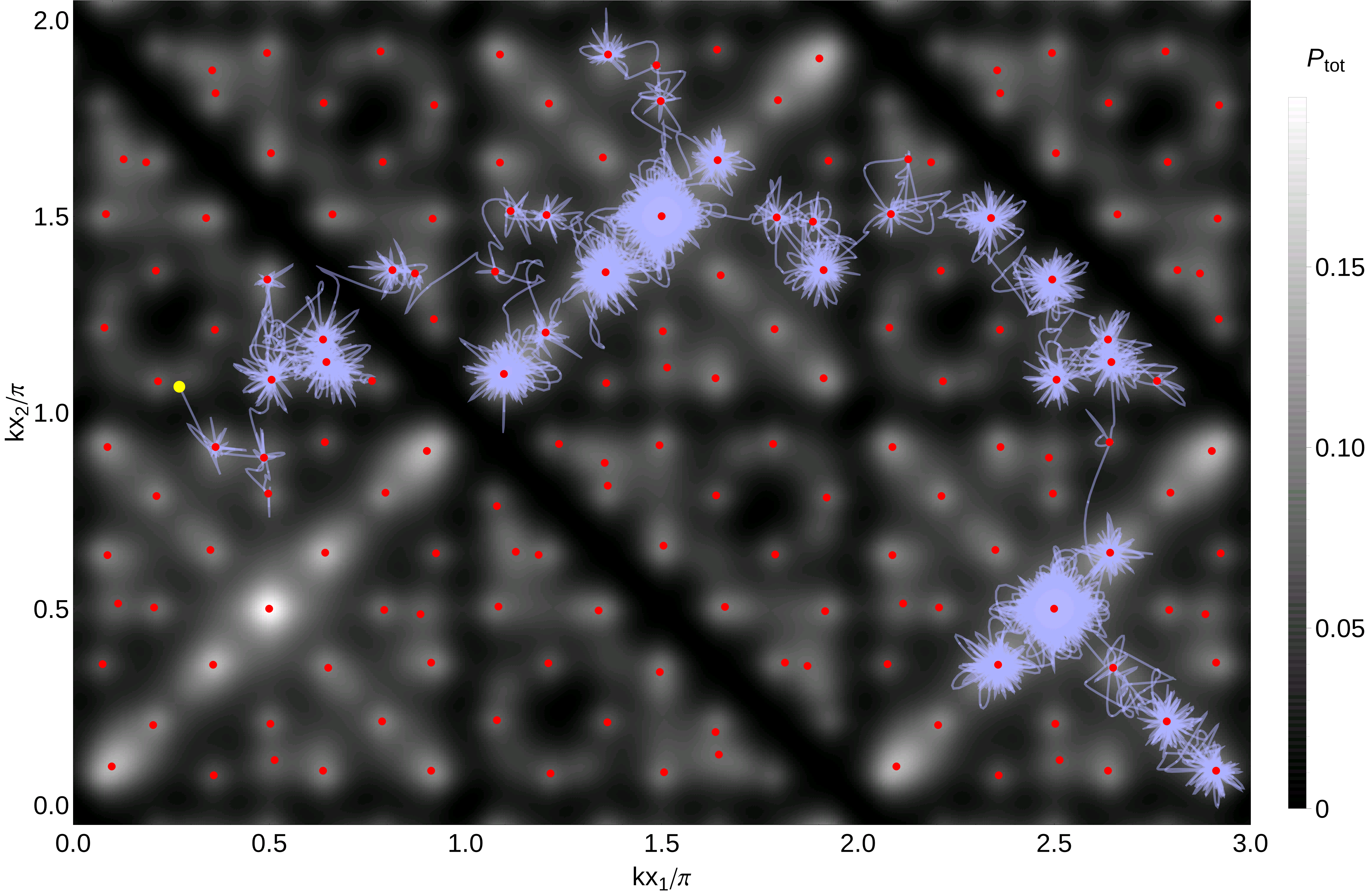}
\caption{Example trajectory of two particles in position space with static illumination and random momentum kicks. The yellow dot indicates the initial position. As in Fig. \ref{photonnumber2}, the density plot in the background represents $P_{tot}$ and the red dots indicate the stable equilibria, both for the overdamped scenario. The illumination pattern is given by $\eta_n = \eta (1,0,1,1,1,0,1)_n$, $n \in \{1,\dots,7\}$ with parameters $\kappa = 10 \omega_R /\pi^2$, $\eta = 2\omega_R/\pi^2$, $U_0 = -5 \omega_R/\pi^2$, $\delta_c = NU_0/2 - 2\kappa$, time step between kicks $\Delta t = \pi^2/5 \omega_R^{-1}$ and friction $\mu = 20 \omega_R/\pi^2$ where $\omega_R = \hbar k^2/2m$.}
\label{staticillumination}
\end{figure}

In any real system damping also adds noise forces on the particles, which can be modeled by random momentum kicks on individual particles as for Brownian motion. To include these kicks, we use the full dynamics (\ref{time_evo}) with a noise term and a linear friction in the momentum equation. Here a stationary distribution is not stable on a long time scale and the system will eventually leave an equilibrium configuration even under a static illumination. In Fig. \ref{staticillumination} we show a corresponding trajectory simulation for two atoms for an extended time. While the particles jump between different equilibrium points the trajectory concentrates close to bright areas of the background picture denoting strong light scattering. The system explores a large volume of configuration space but preferentially stays at points of strong scattering.  

\begin{figure}[h]
\centering
 \includegraphics[width=12cm]{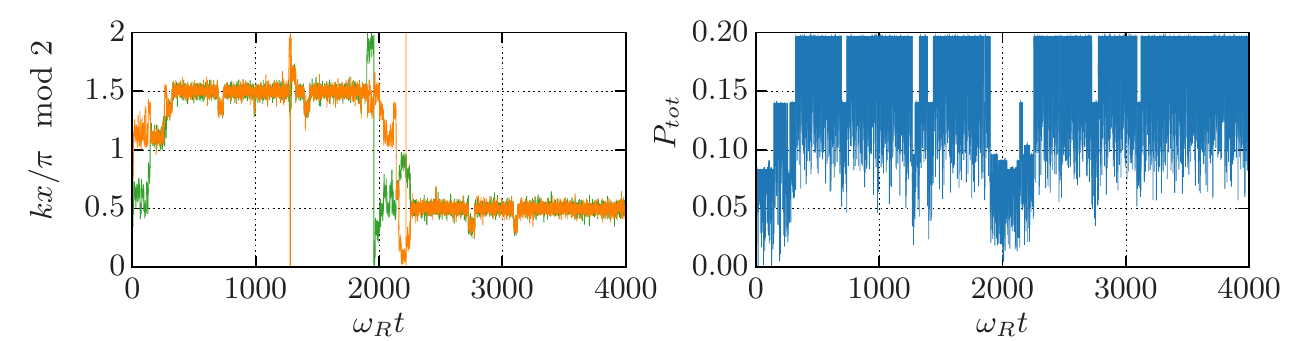}
\caption{Positions of two atoms modulo a wavelength (left) and scattered light $P_{tot}$ (right) as a function of time for the trajectory of Fig. \ref{staticillumination}. Most of the time the atoms are at the points of highest scattering when $k x_{1,2} / \pi$ is either $0.5$ or $1.5$.}
\label{staticillumination-diagrams}
\end{figure}

This effect is exposed more quantitatively in Fig. \ref{staticillumination-diagrams}. On the left we see that often both particles stay in the same well with $k x_{1,2} / \pi$ being either $0.5$ or $1.5$, which corresponds to the points with maximum scattering shown in the right picture. Because available phase space is rather limited for two particles they will often meet each other at the same position where collective scattering is strongest. As the trapping potential is deeper at such points, the time it takes the particles to diffuse out of the corresponding optical well is much larger than for shallow minima with low light scattering. 

\subsection{Time evolution of larger ensembles with randomly varying pump light}

While the above examples nicely illustrate the basic physical mechanism of multicolor self-ordering, more interesting scenarios appear for many higher order modes and larger particle numbers with smaller individual couplings. Since in the corresponding configuration space a huge number of stationary states corresponding to local energy minima exist, it is far less likely for the system to come close to the global optimum and many intermediate configurations appear. In the following we consider hundreds of particles and tens of modes to see to which extend we still get enhanced collective light scattering into several modes. In the classical regime of particle and field dynamics this only requires a moderate increase in computing power. As particle phase space cannot be easily graphically depicted, to present the dynamics in a useful way we select several important collective quantities. It turns out that besides the total light intensity $P_{tot} =  \sum_n |\alpha_n|^2$ scattered into all modes, it is useful to define the sum of all atomic order parameters \cite{Asboth2005self}
\begin{equation}
 \Theta_{tot} = \sum_n |\Theta_n|, \;\;\Theta_n = \frac{1}{N} \sum_j \sin (k_n x_j),
\end{equation}
where $\Theta_n$ quantifies how close the system is to perfect order with respect to the $n$-th mode.

\begin{figure}[h]
 \centering
 \includegraphics[width=13.2cm]{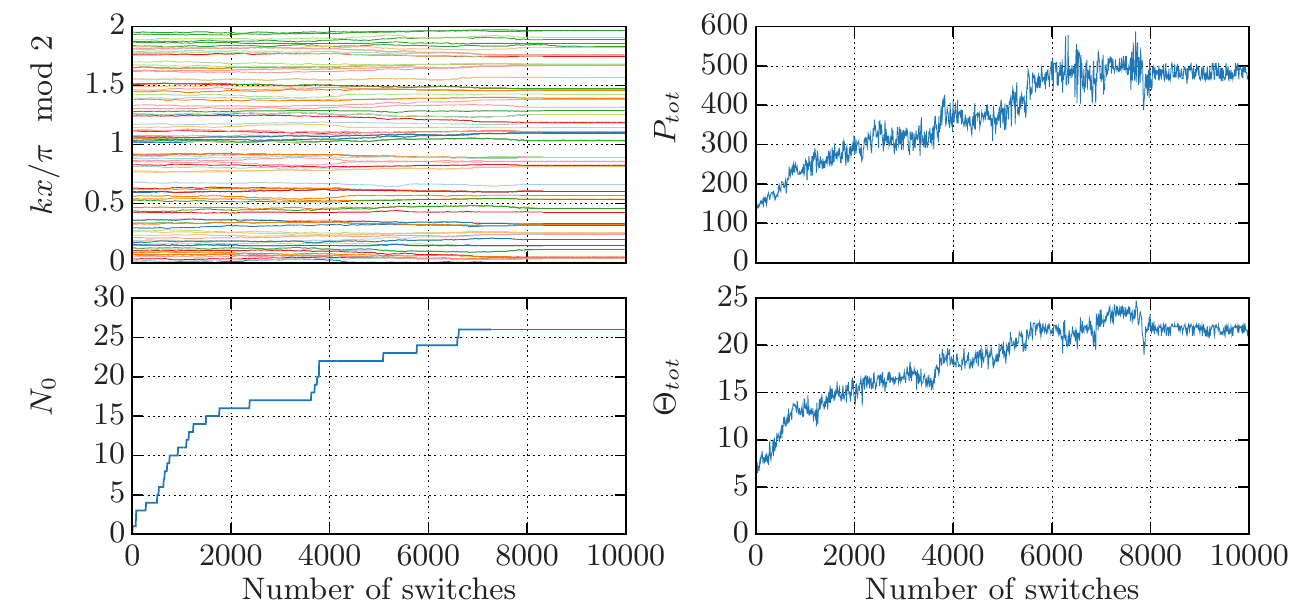}
 \caption{Time evolution of 100 atoms under aperiodically time-varying illumination with about 50 high order modes. The illumination is switched after the atoms find a stationary state. The following quantities are depicted: Atom positions, scattered light $P_{tot}$, total order parameter $\Theta_{tot}$ and the number of clustering particles $N_0$. The parameters are as in Fig. \ref{3dspheres-right}.}
\label{manyparticles}
\end{figure}

Of course for large particle and mode numbers with numerous parameter choices, we can only discuss a few typical cases. As before we assume high field seeking particles and sufficiently red-detuned cavity pumping to avoid heating and nonlinear instabilities. Since we are interested in finding stationary configurations we retreat again to the case of sufficiently strong friction and we ignore noise terms. We choose five illumination patterns consisting of about 50 distinct frequencies $nk$ with equal pump strength $\eta$ chosen out of the set $\{n_1,n_1+\Delta n,\dots,n_1+ 99 \Delta n\}$ with $n_1=1003$ and $\Delta n = 7$.\\
As in chapter \ref{subsec:time-dep-ill}, we switch between the chosen illumination patterns after a prescribed time when a stationary configuration is reached. However, here we randomly choose one of the five patterns at every switch so that no closed loops are attained and the system will evolve non-trivially on a longer time scale.\\
One realization of this dynamics is shown in Fig. \ref{manyparticles}, where we plot the time evolution of the atom positions, the scattered light intensity $P_{tot}$, the order parameter sum $\Theta_{tot}$ and the number of clustering particles $N_0$ (i.e. number of inter-atom distances equal to zero). We observe that $P_{tot}$ and $\Theta_{tot}$ is roughly monotonically increasing demonstrating that the system continuously improves its adaption to the changing illumination. Due to the vast amount of possible configurations, this is a rather slow process continueing at least to 8000 switches. After that the atoms have found areas of configuration space with well adapted positions to all five illumination patterns. Both, the time needed to reach such a state and the resulting light intensity strongly vary for different realizations (i.e. different sequences of illumination patterns) but the qualitative behavior is generally similar. Only, if one of the higher order modes is present in all illumination patterns, the atoms preferentially adapt to this mode with much less light scattered to the others, which results in lower overall scattering in time.\\
Note that the number of particle in clusters $N_0$ grows as $\Theta_{tot}$ ressembling a lower effective particle number with higher individual coupling. In a more realistic model, noise will eventually split existing clusters and prevent lumping of the system. We can also interpret the adaptive ordering as acquisition of a memory of past illumination conditions. We see that when we apply any of the five illumination patterns after a long time evolution, more light will be scattered than in the beginning, where the particles are randomly distributed. Hence, the system memorizes that this illumination has been applied, storing in the prevailing order of the atoms.

\section{Conclusions and outlook}

We have seen that the coupled particle-field evolution of mobile scatterers with multi-frequency illumination in an optical resonator exhibits a wealth of intriguing phenomena beyond simple regular self-ordering. For a proper choice of the detunings and pump powers the particles evolve towards a multitude of different spatial configurations locally minimizing their optical potential energy and at the same time maximizing total light scattering into the cavity. For time varying illumination conditions the system continuously optimizes its light scattering properties and acquires a memory of past conditions. This speeds up adaptation to a new equilibrium when similar conditions reappear, which increases overall scattering efficiency with time. Adding noise and diffusion allows the system to explore larger volumes of configuration space, which results in a configuration diffusion dynamics towards close to optimum scattering conditions for many light frequencies simultaneously. This includes concurrent superradiant scattering into several cavity modes. Hence we can consider the system an adaptive and self-learning light collection system with built-in memory. While implementations with a cold gas in a high-Q cavity would give straightforward possibilities to experimentally study such dynamics, alternative setups using mobile nano-particles in solutions provide equally interesting experimental platforms \cite{douglass2012superdiffusion}.    

\section*{Acknowledgments}

We thank Wolfgang Niedenzu for stimulating discussions and acknowledge support by the Austrian Science Fund via the SFB Foqus project F4013 and I1697-N27.

\end{document}